\documentclass[12pt,tightenlines,eqsecnum,floats,aps,amsmath,amssymb,superscriptaddress,nofootinbib,prd,showpacs]{revtex4}

\usepackage{amssymb}
\usepackage{amsmath}
\usepackage{amsthm}
\usepackage{amsfonts}
\usepackage{graphicx}

\usepackage{subfigure}

\newcommand{\f}{\frac}
\newcommand{\be}{\begin{equation}}
\newcommand{\ee}{\end{equation}}
\newcommand{\bea}{\begin{eqnarray}}
\newcommand{\eea}{\end{eqnarray}}

\begin{document}

\title{Loop Quantum Cosmology: Anisotropy and singularity resolution}

\author{Alejandro Corichi}
\email{corichi@matmor.unam.mx}
\affiliation{Centro de Ciencias Matem\'aticas,
Universidad Nacional Aut\'onoma de M\'exico,
UNAM-Campus Morelia, A. Postal 61-3, Morelia, Michoac\'an 58090,
Mexico}
\affiliation{Center for Fundamental Theory, Institute for Gravitation and the Cosmos,
Pennsylvania State University, University Park PA 16802,
USA}

\author{Asieh Karami}
\email{karami@ifm.umich.mx}
\affiliation{Instituto de F\'{\i}sica y
Matem\'aticas,  Universidad Michoacana de San Nicol\'as de
Hidalgo, Morelia, Michoac\'an, Mexico}
\affiliation{Centro de Ciencias Matem\'aticas,
Universidad Nacional Aut\'onoma de M\'exico,
UNAM-Campus Morelia, A. Postal 61-3, Morelia, Michoac\'an 58090,
Mexico}

\author{Edison Montoya}
\email{edison@ifm.umich.mx}
\affiliation{Instituto de F\'{\i}sica y
Matem\'aticas,  Universidad Michoacana de San Nicol\'as de
Hidalgo, Morelia, Michoac\'an, Mexico}
\affiliation{Centro de Ciencias Matem\'aticas,
Universidad Nacional Aut\'onoma de M\'exico,
UNAM-Campus Morelia, A. Postal 61-3, Morelia, Michoac\'an 58090,
Mexico}

\begin{abstract}
In this contribution we consider the issue of singularity resolution within
loop quantum cosmology (LQC) for different homogeneous models.
We present results of numerical evolutions of effective equations
for both isotropic as well as anisotropic cosmologies, with and without spatial curvature.
To address the issue of singularity resolution we examine the time evolution of
geometrical and curvature invariants that yield information about the semiclassical spacetime
geometry. We discuss generic behavior found for a variety of initial conditions.
Finally, we show that the modifications which come from Loop Quantum Cosmology
imply a non-chaotic effective behavior in the vacuum Bianchi IX model.
\end{abstract}


\maketitle

%%%%%%%%%%%%%%%%%%%%%%%%%%%%%%%%%%%%%%%%%%%%%%%%%%%%
\section{Introduction}

In general relativity (GR) the singularity theorems of Hawking, Penrose and Geroch tell
us that, under reasonable assumptions, singularities are generic. A
spacetime is said to be singular if it is not geodesically complete, which may happen when
some geometrical curvature invariants diverge.
The expectation is that, by quantizing the gravitational degrees of freedom, namely, with
a complete theory unifying gravity and the quantum, the singularities shall be resolved.

Loop quantization (as in Loop Quantum Gravity) of the homogeneous, isotropic and flat
Friedman-Robertson-Walker (FRW) cosmology coupled to a massless scalar field $\phi$, can be exactly solved \cite{slqc}. For that
model it was shown that:
\begin{itemize}
\item The matter density operator $\hat\rho$ has an absolute upper bound and the expansion $\theta$ is also bounded. One
can conclude that curvature scalars do not diverge. This is a signal that a singularity is not present.

\item All states undergo a bounce and with this, the big bang is replaced by a {\it big bounce}.

\item The GR dynamics is recovered as we go away from the Planck scale, this means that we are recovering the original theory that we quantize.

\item Dynamics of semiclassical states are well captured by an effective theory that retains information about the loop quantum geometry.

\item Semiclassical states at late times must have come from semiclassical states before the bounce. That is, semiclassicality is preserved across the bounce \cite{recall,recall-2,CM-1}

\item With all these, one can conclude that the singularities are resolved: the geodesics are inextendible, and are well defined on the other side of the would be big bang.
\end{itemize}

The fact that the {\it effective theory} provides an accurate description of the dynamics at the Planck scale has been  strongly used to explore the anisotropic models. The effective theory is obtained from the quantum
Hamiltonian operator by taking expectation values on appropriately defined states. The thus obtained effective Hamiltonian then generates the dynamics on a classical phase space. The solutions to the
effective theory were shown in \cite{APS2} to  accurately describe the evolution of the expectation value of the observables in the quantum theory when they are considered on semiclassical states. Those results were extended to open, closed and flat FRW models  with and without cosmological constant (see \cite{AS} for a review).
Loop quantum cosmology (LQC) has been extended to the simplest anisotropic
cosmological models, namely Bianchi I, II and IX \cite{Ed-BIX,bianchi1,bianchi2}.
%The quantization of anisotropic models was perform for Bianchi I \ref{}, Bianchi II \ref{} and Bianchi IX \ref{} models.
But in none of these cases has the quantum theory been solved, even numerically.
Then, in order to study these models at the semiclassical level, one generally assumes
that the effective theory reproduces the solutions to the quantum theory when semiclassical states are considered. This is our
working hypothesis, which is well justified by the results in the isotropic cases. It would be
interesting to know whether the evolution of the semiclassical states reproduce the solutions which
we get from the effective theory. From this point of view, the study of the effective theory
can be seen as the first step in this direction.

The new issues to consider in the anisotropic models are: is the bounce generic? We now have anisotropy/Weyl curvature, how does it behave near the singularity/bounce? Can we have different kind of bounces, say, dominated by shear $\sigma^2$? Are the geometric scalars such as the expansion
$\theta$, the shear $\sigma^2$ and density $\rho$
absolutely bounded? The goal of this contribution is to answer these questions using the effective theory for Bianchi I which has anisotropies,
Bianchi II that has anisotropies and spatial curvature and Bianchi IX which has all the features of Bianchi I, II and is,
furthermore,  spatially compact. Even  more, the Bianchi IX model has a
%, inverse triad corrections
non trivial classical limit, in the sense that, vacuum Bianchi IX is chaotic in the classical theory
and behaves like Bianchi I  with Bianchi II transitions as one approaches the singularity.
In all the cases that we shall consider, the matter content shall consist of a massless scalar field playing the role of internal time. In the case of Bianchi IX we shall also consider the vacuum limit.

%%%%%%%%%%%%%%%%%%%%%%%%%%%%%%%%%%%%%%%%%%%%%%%%%%%%%%
\section{Preliminaries}

In this section we briefly review the quantization of some cosmological models which include $k$=0 and $k$=1 FRW and Bianchi I, II and IX models by using loop quantum gravity methods.
Let us consider the spacetime as $M=\Sigma\times\mathbb R$ where $\Sigma$ is a spatial 3-manifold which can be identified by the symmetry group of the chosen model and is endowed with a fiducial metric ${}^oq_{ab}$ and associated fixed fiducial basis of 1-forms ${}^o\omega_a^i$ and vectors ${}^oe_i^a$. If $\Sigma$ is non-compact then
we fix a fiducial cell, $\mathcal V$, adapted to the fiducial triads with finite volume $V_o$. We also define $L_i$ which is the length of the $i$th side of the cell along ${}^oe_i$ and $V_o=L_1L_2L_3$. We choose for compact $\Sigma$, $L_i=V_o^{1/3}$ with $i=1,2,3$.

In general relativity, the gravitational phase space consists of pairs $(A_a^i,E_i^a)$ on $\Sigma$ where $A_a^i$ is a SU(2) connection and $E_i^a$ is a densitized triad of weight 1. Since all of the models in which we are interested are homogeneous and, if we restrict ourselves to diagonal metrics, one can fix the gauge in such a way that
$A_a^i$ has 3 independent components, $c^i$, and $E_i^a$ has 3 independent components, $p_i$,
\begin{equation}
A_a^i=\frac{c^i}{L_i}{}^o\omega_a^i\ \ \ \textrm{and}\ \ \ E_i^a=\frac{p_iL_i}{V_o}\sqrt{{}^oq}{}\ ^oe_i^a
\end{equation}
where $p_i$ in terms of the scale factors $a_i$ are $|p_i|=L_iL_ja_ja_k$ ($i\neq j\neq k$). Note that in isotropic cases, each of phase space variables has only one independent component.
Using $(c^i,p_i)$ for anisotropic models,
the Poisson brackets can be expressed as $\{c^i,p_j\}=8\pi G\gamma\delta_j^i$ and for isotropic models, the Poisson
bracket is $\{c,p\}=8\pi G\gamma/3$ where $\gamma$ is Barbero-Immirizi parameter.
With this choice of variables and gauge fixing, the Gauss and diffeomorphism constraints are satisfied and the only constraint is the Hamiltonian constraint
\begin{equation}\label{FHC}
\mathcal C_H=\int_\mathcal V N\left[-\frac{\epsilon^{ij}_{\ k}E_i^aE_j^b}{16\pi G\gamma^2\sqrt{|q|}}\left(
F_{ab}^k-(1+\gamma^2)\Omega_{ab}^k\right)+\mathcal H_{\rm matter}\right]\textrm{d}^3x \, ,
\end{equation}
with $N$ the lapse function, $\mathcal H_{matter}=\rho V$ and $\Omega_{ab}$ the curvature of spin connection $\Gamma_a^i$ compatible with the triads.\\
To construct the quantum kinematics, we have to select a set of elementary observables such that their associated operators are unambiguous. In loop quantum gravity they are the holonomies $h_e$ defined by the connection $A_a^i$ along edges $e$ and the fluxes of the densitized triad $E_i^a$ across surfaces.
For our homogeneous models we choose holonomies and $p_i$.
To have the corresponding constraint operator, one needs to express it in terms of the chosen phase space functions $h_e$ and $p_i$.
The first term, $\epsilon^{ij}_{\ k}E_i^aE_j^b/\sqrt{|q|}$, as in loop quantum gravity, can be treated by using Thiemann's strategy \cite{TT}.
\begin{equation}
\epsilon_{ijk}\frac{E^{ai}E^{bj}}{\sqrt{|q|}}=\sum_i\frac{1}{2\pi\gamma G\mu}\ ^o\epsilon^{abc}\ ^o\omega_c^i\textrm{Tr}(h_i^{(\mu)}\{h_i^{(\mu)-1},V\}\tau_k)
\label{ths}
\end{equation}
where $h_i^{(\mu)}$ is the holonomy along the edge parallel to $i$-th vector basis with length $\mu$ and $V$ is the volume, which is equal to $\sqrt{|p_1p_2p_3|}$. Note that $\mu$ is arbitrary.
Now, to define an operator related to the first term of Eq.(\ref{FHC}), we can use the right hand side of
Eq.(\ref{ths}) and replace Poisson brackets with commutators.
To find an operator related to the curvature $F_{ab}^k$, for isotropic models and Bianchi I, one can consider a square
$\square_{ij}$ in the $i-j$ plane which is spanned by two of the fiducial triads (for the closed isotropic model since
triads do not commute, to define this plane we use a triad and a right invariant vector ${}^o\xi_i^a$),
with each of its sides having length $\mu_i^\prime$. Therefore, $F_{ab}^k$ is given by
\begin{equation}
F_{ab}^k=2\lim_{Area_\square\rightarrow 0}\epsilon_{ij}^{\ \ k}\textrm{Tr}\bigg(\frac{h_{\square_{ij}}^{\mu^\prime}-\mathbb I}{\mu^\prime_i\mu^\prime_j}\tau^k\bigg){}^o\omega_a^i{}^o\omega_b^j \, .
\label{fs}
\end{equation}
Since in loop quantum gravity, the area operator does not have a zero eigenvalue, one can take the limit of Eq.(\ref{fs})
to the point where the area is
equal to the smallest eigenvalue of the area operator, $\lambda^2 =4 \sqrt{3} \pi \gamma l_{p}^2$,
instead of zero. Then, $\mu_i^\prime a_i=\lambda$. We take  $\mu_i^\prime=\bar\mu_i L_i$ where $\bar\mu_i$ is a dimensionless parameter and, by previous considerations,  is equal to $\bar\mu_i=\lambda\sqrt{|p_i|}/\sqrt{|p_jp_k|}$ ($i\neq j\neq k$).

For Bianchi II and IX, we cannot use this method because the resulting operator is not almost periodic, therefore
we express the connection $A_a^i$ in terms of holonomies and then use the standard definition of curvature $F_{ab}^k$.
The operators corresponding to the connection are given by \cite{bianchi2}
\begin{equation}
\hat c_i=\widehat{\frac{\sin\bar\mu_ic_i}{\bar\mu_i}}\, .
\end{equation}
Note that using this quantization method for flat FRW and Bianchi I models, one has the same result as the direct quantization of curvature $F_{ab}^k$, but for closed FRW it leads to a different quantum theory which is more compatible with the isotropic limit of Bianchi IX. We call the first method of quantization {\it curvature based quantization} and the second one {\it connection based quantization}.
In Bianchi II and Bianchi IX models the terms related to the curvatures,
$F_{ab}^k$ and $\Omega_{ab}^k$, contain some negative powers of $p_i$ which are not well defined operators. To solve this problem we use the same idea as Thiemann's strategy.
\begin{equation}
|p_i|^{(\ell-1)/2}=-\frac{\sqrt{|p_i|}L_i}{4\pi G\gamma j(j+1)\tilde\mu_i\ell}\textrm{Tr}(\tau_i h_i^{(\tilde\mu_i)}\{h_i^{(\tilde\mu_i)-1},|p_i|^{\ell/2}\}) \, ,
\label{np}
\end{equation}
where $\tilde\mu_i$ is the length of a curve, $\ell \in (0,1)$ and $j\in \frac{1}{2}\mathbb{N}$ is for the representation.
Therefore, for these three different operators we have three different curve lengths ($\mu,\mu^\prime,\tilde\mu$) where $\mu$ and
$\tilde\mu$ can be some arbitrary functions of $p_i$, so for simplicity
we can choose all of them to be equal to $\mu^\prime$. On the other hand we have another free parameter in the definition of
negative powers of $p_i$ where, for simplicity, we take $j=1/2$. Since the largest negative power of $p_i$
which appears in the constraint is $-1/4$, we will take $\ell=1/2$ and obtain it directly from Eq.(\ref{np}),
and after that we express the other negative powers by them.
The eigenvalues for the operator $\widehat{|p_i|^{-1/4}}$ are given by
\begin{equation}
J_i(V,p_1,p_2,p_3)=\frac{h(V)}{V_c}\prod_{j\neq i}p_j^{1/4}\, ,
\end{equation}
with
\begin{equation}
h(V)=\sqrt{V+V_c}-\sqrt{|V-V_c|},\,\, \textrm{ and } \,\,\, V_c=2\pi\gamma\lambda\ell_p^2.
\end{equation}

By using these results and choosing some factor ordering, we can construct the total constraint operator. Note that  different choices of factor
ordering will yield different operators, but the main results will remain almost the same. By solving the constraint equation
$\hat{\mathcal C}_H\cdot\Psi=0$, we can obtain the physical states and the physical Hilbert space $\mathcal H_{\rm phys}$.
As a final step, one would need to identify the physical observables, that in our case would correspond to relational observables as functions of the internal time $\phi$.
\\
To test singularity resolution we will study some geometric observables: expansion $\theta$, shear $\sigma^2$,
curvature scalars and also volume of the universe $V$ and matter density $\rho$, as relational observables in terms of $\phi$.
%, the massless scalar field.
\\
Since working with full quantum theories of the models is difficult and, as shown in \cite{APS2}
for some models, the behavior of the effective
or semiclassical equations, which are `classical' equations with some quantum corrections, are good approximations to the numerical quantum evolutions even near the Planck scale, we will work with the effective equations.

%%%%%%%%%%%%%%%%%%%%%%%%%%%%%%%%%%%%%%%%%%%%%%%%%
\section{Effective Theories}

%%%%%%%%%%%%%%%%%%%%%%%%%%%%%%%%%%%%%%%%%%%%%%
The purpose of this section is to present the effective theories obtained for the different models considered.
It has four parts. In the first one, we consider the flat isotropic model and discuss its main features. In the second part
we recall the $k$=1 FRW model for both quantizations available. The third part deals with the Bianchi I and II models.
The Bianch IX model is considered in the last part.

\subsection{Isotropic Flat Model}

In the FRW model with $k$=0, the {\it effective} Hamiltonian is given by
\begin{equation}
\mathcal{H}_{k=0}=\frac{3}{8\pi G\gamma^2\lambda^2}\,V^2 \,\sin(\lambda\beta)^2 -\frac{p_\phi^2}{2}\, ,
\end{equation}
where $p_\phi$ is the momentum of the field, $V$ is the volume and $\beta$ its conjugate variable.
They are related to the $c$ and $p$ variables by the equations
$V=p^{3/2}$, $\beta = c/\sqrt{p}$ and satisfy the Poisson bracket $\{\beta,V\}=4\pi G \gamma$ and $\{\phi,p_\phi\}=1$.
It was shown \cite{APS2} that the dynamics of semiclassical states are well captured by the effective Friedman equation
\[ H^2 = \frac{8\pi G}{3}\rho \left( 1- \frac{\rho}{\rho_{\rm crit}} \right) \, , \]
with $H=\dot{V}/3V$ the Hubble parameter and $\rho=p_\phi^2/2V^2$ the matter density.
The GR dynamics is recovered as we go away from the Planck scale, that is, for densities
$\rho < \rho_{\rm crit}/10$,
with the critical density given by,
$\rho_{\rm crit} = 3/8\pi G \gamma^2\lambda^2\approx 0.41\rho_p$ where $\rho_p$ is the Planck density.
Thus, all trajectories undergo a bounce where the density reaches precisely its critical value $\rho_{\rm crit}$.

%%%%%%%%%%%%%%%%%%%%%%%%%%%%%%%%%%%%%%%%%%%%%%
\subsection{Isotropic Closed Model}

Now, for the isotropic closed model,
as we discussed in previous section, there are two different quantum theories depending on the two different methods of quantization of the curvature $F_{ab}^k$. The Hamiltonians are \cite{CK-2},
%\vspace{-11pt}
\begin{align}
\label{k1}
&\mathcal{H}_{\rm k=1}^{(1)}=\frac{3V^2}{8\pi G\gamma^2 \lambda^2}
[\sin^2(\lambda\beta-D)-\sin^2D+(1+\gamma^2)D^2]-\frac{p_\phi^2}{2}\approx 0 \\
&\mathcal{H}_{\rm k=1}^{(2)}=\frac{3V^2}{8\pi G\gamma^2 \lambda^2}
[\sin^2\lambda\beta-2D\sin\lambda\beta+(1+\gamma^2)D^2]-\frac{p_\phi^2}{2}\approx 0 \, ,
\end{align}
where $D=\lambda\vartheta/V^{-1/3}$ and $\vartheta=(2\pi^2)^{1/3}$.
From the Hamiltonians, the corresponding quantum corrected Friedmann equations are
\begin{align}
\label{H-k1}
&H^2 = \frac{8\pi G}{3}(\rho-\rho\prime) \left( 1- \frac{\rho-\rho\prime}{\rho_{\rm crit}} \right) \\
&H^2 = \frac{8\pi G}{3}(\rho-\rho_1) \left( 1- \frac{\rho-\rho_2}{\rho_{\rm crit}} \right) \\
\nonumber
\end{align}
where $\rho\prime=\rho_{\rm crit}[(1+\gamma^2)D^2-\sin^2D]$, $\rho_1=\rho_{\rm crit}\gamma^2D^2$
and $\rho_2=\rho_{\rm crit}D[(1+\gamma^2)D-2\sin\lambda\beta]$.
Since for both effective theories there are some geometric observables which are not absolutely bounded,
%since to calculate the equations we take the expectation values up to their leading term,
we go further and use more corrections which come from the inverse triad term in the full theory, to see if the
unboundedness of those observables are generic, or whether they improve by adding more corrections.
Therefore, the Hamiltonian constraints change to
\begin{align}
\label{c-k1}
&\mathcal{H}_{\rm k=1}^{(1)}=\frac{3A(V)V}{8\pi G\gamma^2 \lambda^2}
[\sin^2(\lambda\beta-D)-\sin^2D+(1+\gamma^2)D^2]-\frac{p_\phi^2}{2}\approx 0 \\
\label{c-k11}
&\mathcal{H}_{\rm k=1}^{(2)}=\frac{3A(V)V}{8\pi G\gamma^2 \lambda^2}
[\sin^2\lambda\beta-2D\sin\lambda\beta+(1+\gamma^2)D^2]-\frac{p_\phi^2}{2}\approx 0  \, ,
\end{align}
with
\begin{equation}
A(V)=\frac{1}{2V_c}(V+V_c-|V-V_c|)=
\left\{\begin{array}{lr}V/V_c & V< V_c\\1 & V\geq V_c\end{array}\right.
\end{equation}
is a correction term which comes from the operator $\epsilon_k^{ij} E_i^aE_j^b/\sqrt{|q|}$.

%%%%%%%%%%%%%%%%%%%%%%%%%%%%%%%%%%%%%%%%%%%%%%
\subsection{Bianchi I and II}

The effective Hamiltonian for Bianchi I and II can be written in a single expression,
\begin{align*}
\label{H-BII}
\mathcal{H}_{\rm BII} & = \f{p_1p_2p_3}{8\pi G\gamma^2\lambda^2}
\left[\f{}{}\sin\bar\mu_1c_1\sin\bar\mu_2c_2+\sin\bar\mu_2c_2
\sin\bar\mu_3c_3+\sin\bar\mu_3c_3\sin\bar\mu_1c_1\right] \nonumber\\
& \quad + \f{1}{8\pi G\gamma^2}
\Bigg[\f{\alpha(p_2p_3)^{3/2}}{\lambda\sqrt{p_1}}\sin\bar\mu_1c_1
-(1+\gamma^2)\left(\f{\alpha p_2p_3}{2p_1}\right)^2 \Bigg] - \f{p_\phi^2}{2} \approx 0 \, .
\end{align*}
The parameter $\alpha$ allows us to distinguish between Bianchi I ($\alpha=0$) and Bianchi II ($\alpha = 1$).
This Hamiltonian together with the Poisson Brackets $\{c^i,p_j\}=8\pi G\gamma\delta_j^i$ and $\{\phi,p_\phi\}=1$
gives the effective equations of motion. % ($\dot\phi,\dot p_\phi,\dot c^i,\dot p_i$).
It has been shown that the density, the expansion and the shear are absolutely bounded for the case of Bianchi I
\cite{Corichi-geometric, Singh-BI}. The case of Bianch II is more involved, as we shall see later on.

%%%%%%%%%%%%%%%%%%%%%%%%%%%%%%%%%%%%%%%%%%%%%%
\subsection{Bianchi IX}
In the previous cases, the effective Hamiltonian were found by choosing the lapse function as $N=V$.
But now in Bianchi IX, we choose $N=1$. This will allow us to
 include more inverse triad corrections.
Then, the effective Hamiltonian is given by \cite{CK-3}
\bea
\label{H-BIX}
\mathcal{H}_{\rm BIX}&=&-\frac{V^4A(V)h^6(V)}{8\pi GV_c^6\gamma^2\lambda^{2}}\bigg(\sin\bar\mu_1c_1\sin\bar\mu_2c_2+\sin\bar\mu_1c_1\sin\bar\mu_3c_3
+\sin\bar\mu_2c_2\sin\bar\mu_3c_3\bigg)\nonumber\\
& &+\frac{\vartheta A(V)h^4(V)}{4\pi GV_c^4\gamma^2\lambda}\bigg(p_1^2p_2^2\sin\bar\mu_3c_3
+p_2^2p_3^2\sin\bar\mu_1c_1 +p_1^2p_3^2\sin\bar\mu_2c_2\bigg)\nonumber\\
& &-\frac{\vartheta^2(1+\gamma^2)A(V)h^4(V)}{8\pi GV_c^4\gamma^2}\bigg(2V[p_1^2+p_2^2+p_3^2]
-\bigg[(p_1p_2)^{4}+(p_1p_3)^{4}+(p_2p_3)^{4}\bigg]\frac{h^6(V)}{V_c^6}\bigg)\nonumber\\
& &+\f{h^6(V)V^2}{2V^6_c}p_\phi^2 \approx 0
\eea

%%%%%%%%%%%%%%%%%%%%%%%%%%%%%%%%%%%%%%%%%%%%%%%%
\section{Effective Vacuum Bianchi IX}
In this section we study the vaccum Bianchi IX due to its non trivial classical behavior, in the sense that
near to the classical singularity it presents a
chaotic behavior described by the Belinski, Khalatnikov and Lifshitz (BKL) scenario. This case can be considered within our analysis because it is
included in the Bianchi IX case with massless scalar field when we take the  $p_\phi = 0$  limit in Eq.(\ref{H-BIX}).
To study the BKL scenario within the effective theory we use the same idea given by Misner \cite{MI, Mb}, i.e.,
we study the potential term in the Hamiltonian constraint. Due
to the new features introduced by the effective theory, to study the potential term is not enough. Therefore,
we analyze also the density in order to determine how the BKL scenario changes.

%%%%%%%%%%%%%%%%%%%%%%%%%%%%%%%%%%%%%%%%%%%%%%%%
\subsection{The Effective Potential}
It is helpful to use the potential term of the constraint to study the solutions. The classical potential which comes from the spin connection's curvature in the classical constraint, in terms of Misner variables is \cite{MI}
\begin{equation}
W=\frac{1}{2} e^{-4\Omega}\bigg(e^{-4\beta_+}-4e^{-\beta_+}\cosh{\sqrt 3\beta_-}+2e^{-2\beta_+}[\cosh{2\sqrt 3\beta_-}-1]\bigg) \, ,
\end{equation}
where $\Omega=-\frac{1}{3}\log V$ and the anisotropies $\beta_\pm$ are defined via
\be
a_1=e^{-\Omega+(\beta_++\sqrt 3\beta_-)/2},\quad a_2=e^{-\Omega+(\beta_+-\sqrt 3\beta_-)/2}, \quad a_3=e^{-\Omega-\beta_+}.
\ee
Since the $\Omega$ dependence factorizes, one can obtain an anisotropy potential $\mathcal V(\beta_+,\beta_-)$ which exhibits exponential walls for large
anisotropies. The universe can be seen as a particle moving in such a potential ($W$) that presents reflections at the walls.
An infinite number of these reflections implies that the system behaves chaotically.
When the volume becomes small, the quantum effects become important
and one should work with the full quantum theory, but one can use the effective equations to have a qualitative view of what happens near the classical singularity. From the effective Hamiltonian Eq.(\ref{H-BIX}), the modified potential can be derived as a function of $p_i$
\begin{equation}
W_{\rm eff}=-\frac{V^2 A(V)h^4(V)}{V_c^4}\bigg(p_1^2+p_2^2+p_3^2-\bigg[(p_1p_2)^{4}+(p_1p_3)^{4}+(p_2p_3)^{4}\bigg]\frac{h^6(V)}{2VV_c^6}\bigg) \, .
\end{equation}
For a simple case, when $\beta_-=0$ and $\beta_+\rightarrow -\infty$, the classical potential is
$W(\beta_+,\Omega)\sim\frac{1}{2} e^{-4\Omega-4\beta_+}$. If we rewrite the modified potential in terms of Misner
variables, we can see that in this limit, the modified potential behaves as $W_{\rm eff}\sim\frac{1}{2V_c^9}e^{-52\Omega-4\beta_+}$,
where the $\beta_+$-dependency of both classical and modified potential are the same.
Thus, we have an infinite wall for the modified potential, too
(See Fig.\ref{pot}, left). However, for small volumes, the modified potential can be negative at some points.

One should note that the kinetic term is also modified. However, this modification does not change the qualitative behavior, 
because the kinetic term is always finite and different form zero for a finite time.

%%%%%%%%%%%%%%%%%%%%%%%%%%%%%%%%%%%%%%%%%%%%%%%%%%
\subsection{Density}
\begin{figure}
        \centering
        \begin{subfigure}
                \centering
                \includegraphics[width=0.5\textwidth, height=0.45\linewidth]{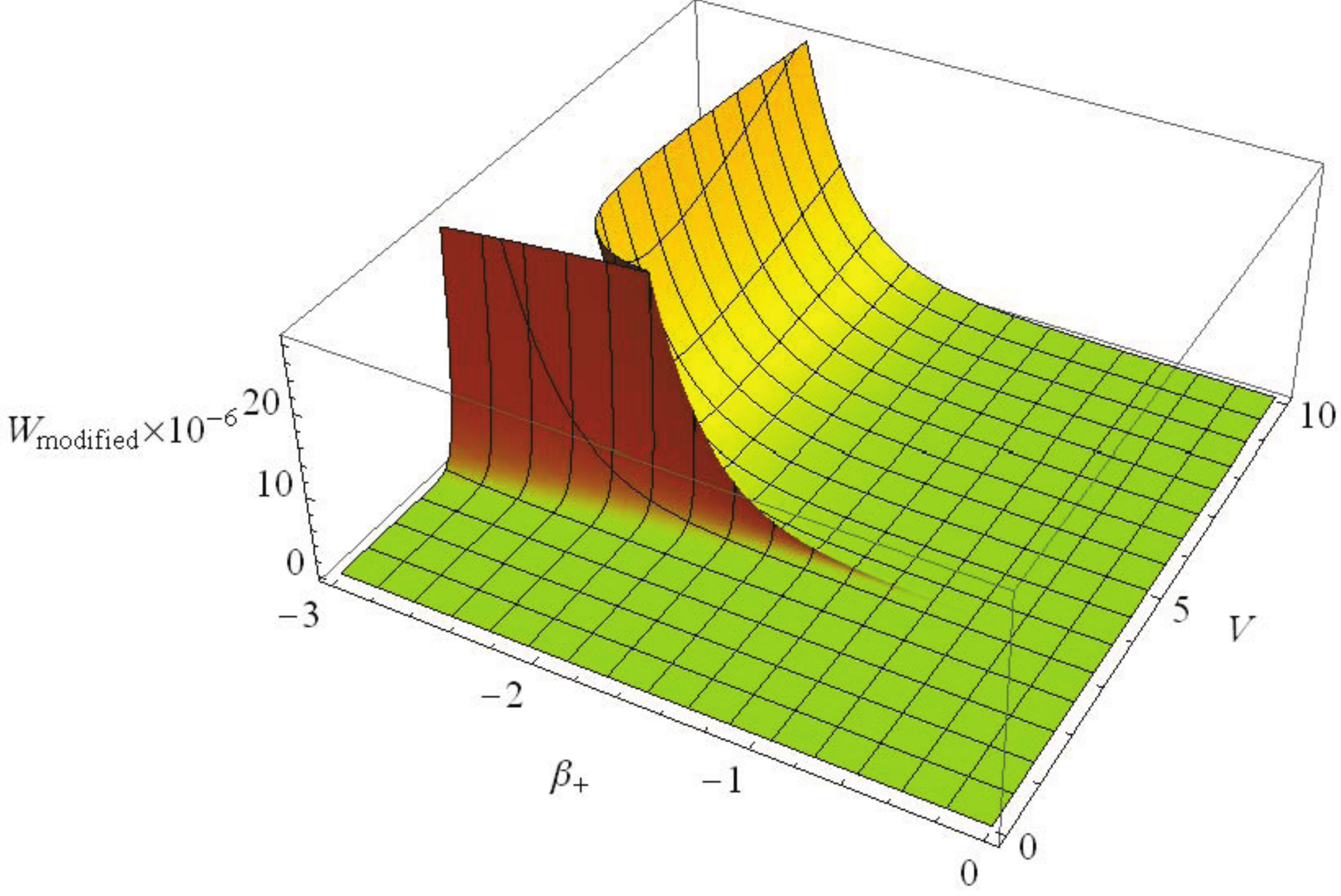}
        \end{subfigure}
     \begin{subfigure}
                \centering
                \includegraphics[width=0.45\textwidth, height=0.45\linewidth]{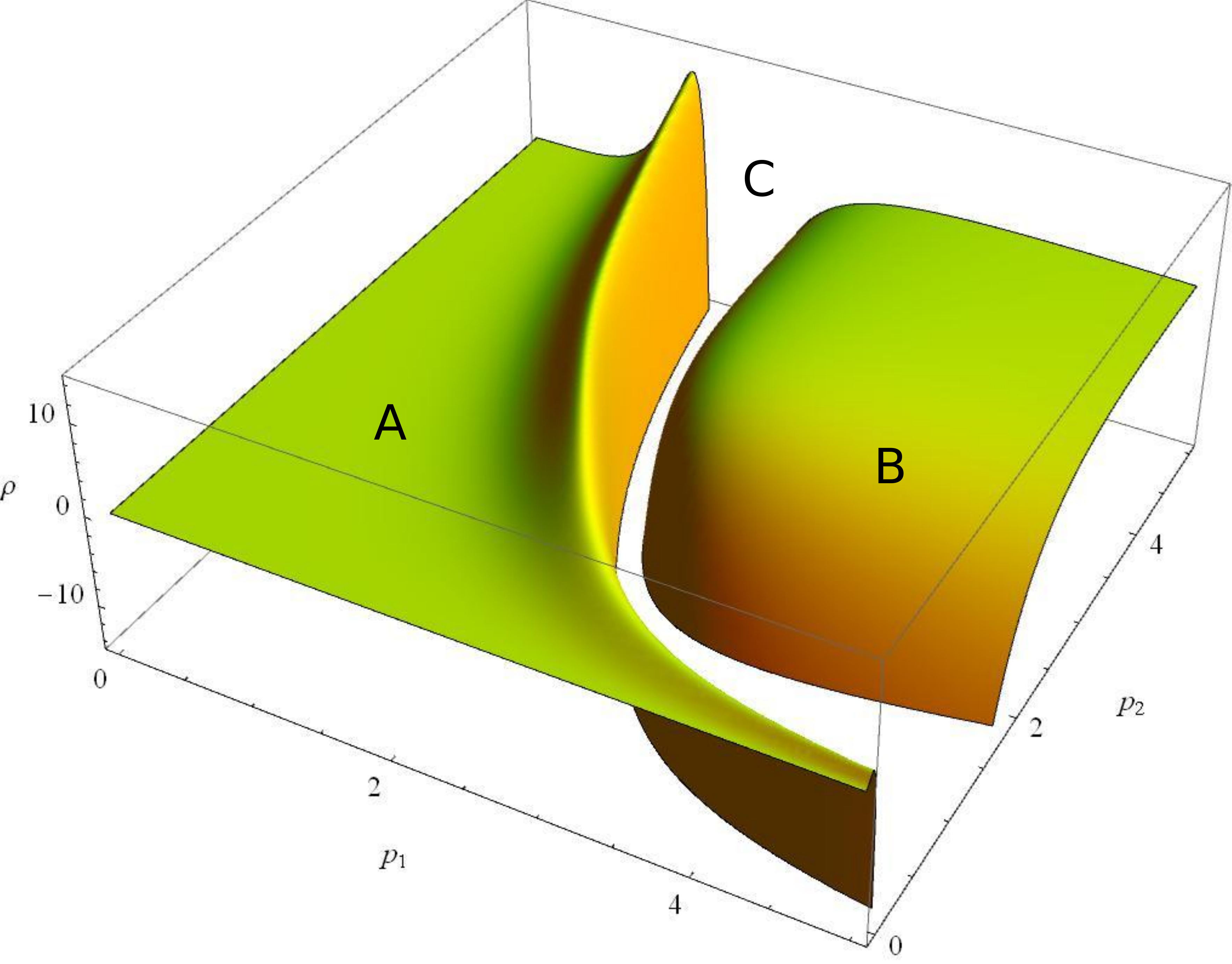}
        \end{subfigure}
\caption{Left, modified potential when $\beta_-=0$. Right, maximum allowed density vs. $p_1$ and $p_2$ where $p_3=p_1$. Both in Planck units.}
\label{dens} \label{pot}
\end{figure}

In the general case, as it can be seen in Fig.\ref{dens} (right), the maximum allowed density which arises from the modified Hamiltonian, has two distinct
disconnected regions with positive values. If we impose the weak energy condition and start the evolution within one region, 
the universe cannot reach the other region. These two regions have different dynamics.
To study the vacuum Bianchi IX, we start from large volumes which lie in region B of Fig.\ref{dens} (right) and, as we go to smaller volumes we cannot reach  zero volume because `crossing' to region A is not allowed. Therefore, there is a smallest reachable volume in region B and, since very large anisotropies are not allowed near this smallest volume, and the modified potential is not too large there, then we have, at most, finite oscillations before reaching the bounce. On the other hand, in the internal region A, the anisotropies are very large when some of the $p_i$ are very small, and then the volume of the universe cannot be large enough to start the evolution from there.

%\vspace{-11pt}
%%%%%%%%%%%%%%%%%%%%%%%%%%%%%%%%%%%%%%%%%%%%%%%%%%%
\section{Results}

We will now compare the results of the effective theories for the isotropic FRW $k$=0 and $k$=1,
diagonal Bianchi I, II and IX. For all of them,
the  matter content consists of a massless scalar field satisfying the Klein-Gordon equation.
A good starting point to compare the results
is to answer the questions that we posed in the introduction,

\begin{itemize}

\item Is the bounce generic? Yes. All solutions have a bounce.
In other words, singularities are resolved.
%i.e., they provide generic paradigm of singularity resolution.
In the closed FRW and the Bianchi IX model, there is an infinite number of bounces and recollapses due to the compactness of the spatial manifold.

\item How does anisotropy/Weyl curvature behave near the bounce?
For Bianchi I and II the shear and the curvature scalars, far from the bounce, are monotonic and approach their
classical values, but when they reach the region near the bounce they behave differently. In Bianchi I, they present one maximum which occurs
at the bounce. In Bianchi II, they exhibit a richer behavior, because now they can be zero at the bounce or near to it, see Fig. \ref{fig-b2},
and have more than one maximum
(for the shear there are up to 4 maxima, shown in Fig. \ref{fig-b2}, and for the scalar curvature up to 2 maxima, see \cite{CM-bianchi2}).
In Bianchi IX, if we restrict the analysis to one of the infinite number of bounces, it can be shown that anisotropy and curvature
behave as in the Bianchi I or II cases. Of currect research is whether there are new behaviors \cite{CKM}.
When we consider more than one bounce in the Bianchi IX case, there appear new interesting behaviors,
shown in Fig. \ref{fig-b9} and Fig. \ref{fig-b9-2}.

%%%%%%%%%%%%%%%%%%%%%%%%%%%%%%%%%%%%%%%%%%%%%%%%%%%%%%%%%
\begin{figure}
        \centering
        \begin{subfigure}
                \centering
                \includegraphics[width=0.47\textwidth]{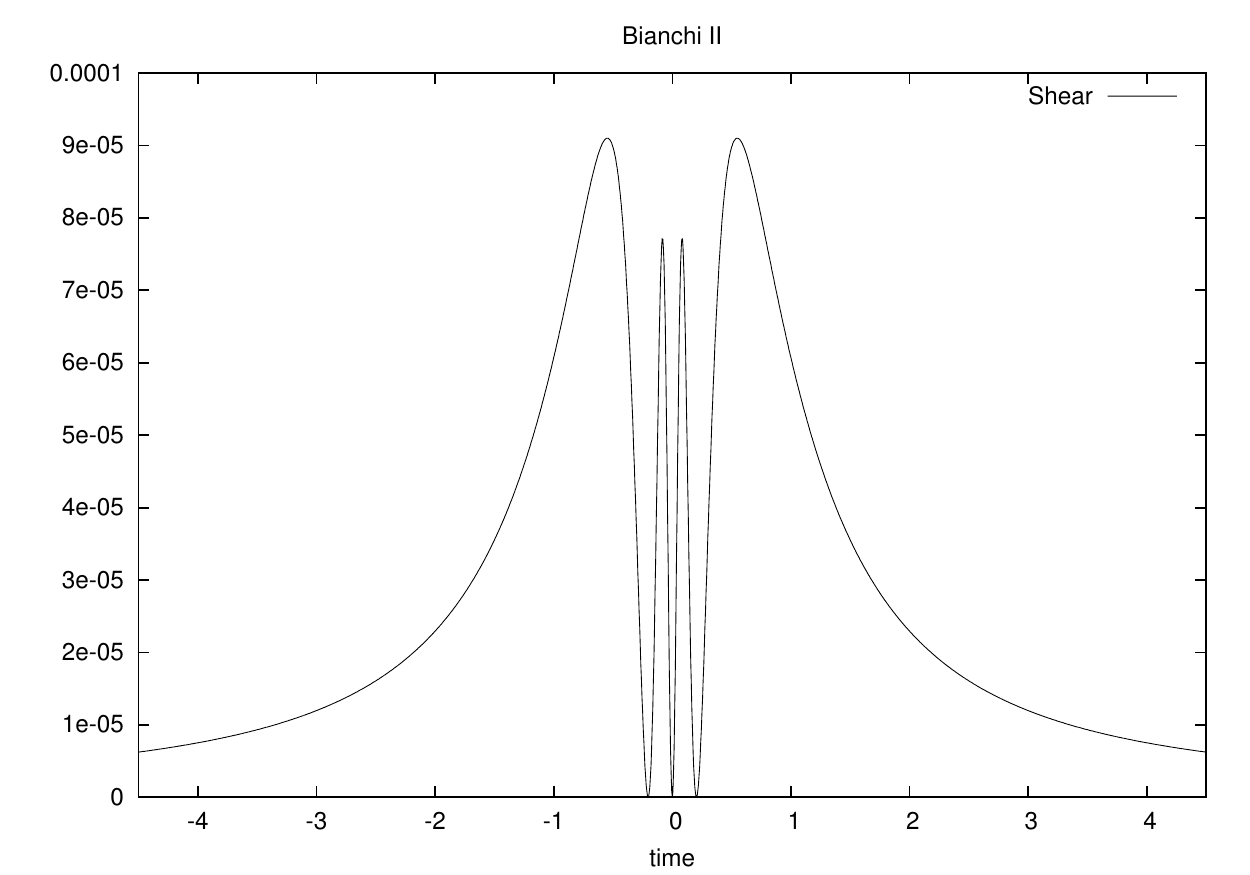}
        \end{subfigure}
     \begin{subfigure}
                \centering
                \includegraphics[width=0.47\textwidth ]{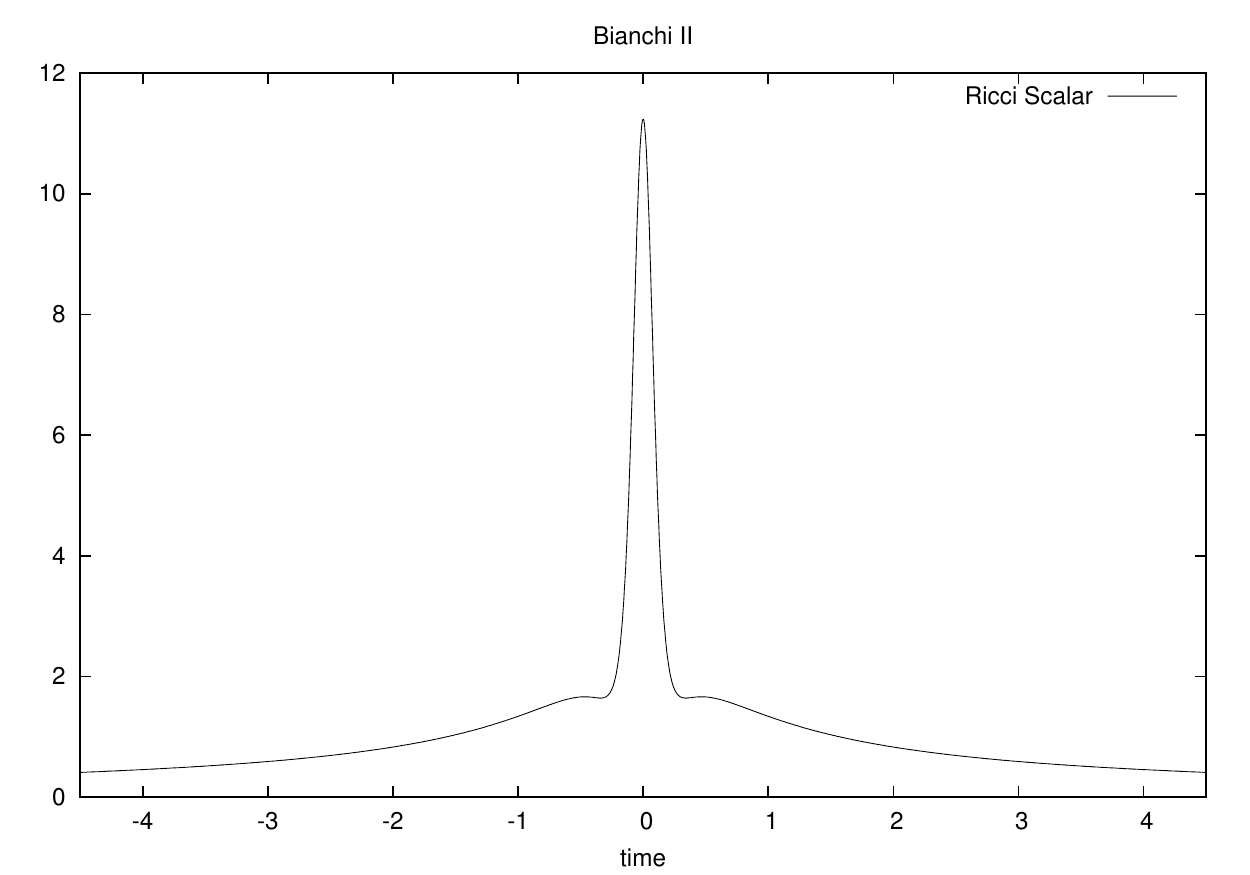}
        \end{subfigure}
\caption{Evolution of shear $\sigma^2$ and Ricci scalar $R$ for Bianchi II. The initial conditions at $t=0$ are: $\bar\mu_ic_i=\pi/2$ and $p_i=1000$. }
\label{fig-b2}
\end{figure}

%%%%%%%%%%%%%%%%%%%%%%%%%%%%%%%%%%%%%%%%%%%%%%%%%%%%%%%%%
\begin{figure}
        \centering
        \begin{subfigure}
                \centering
                \includegraphics[width=0.47\textwidth]{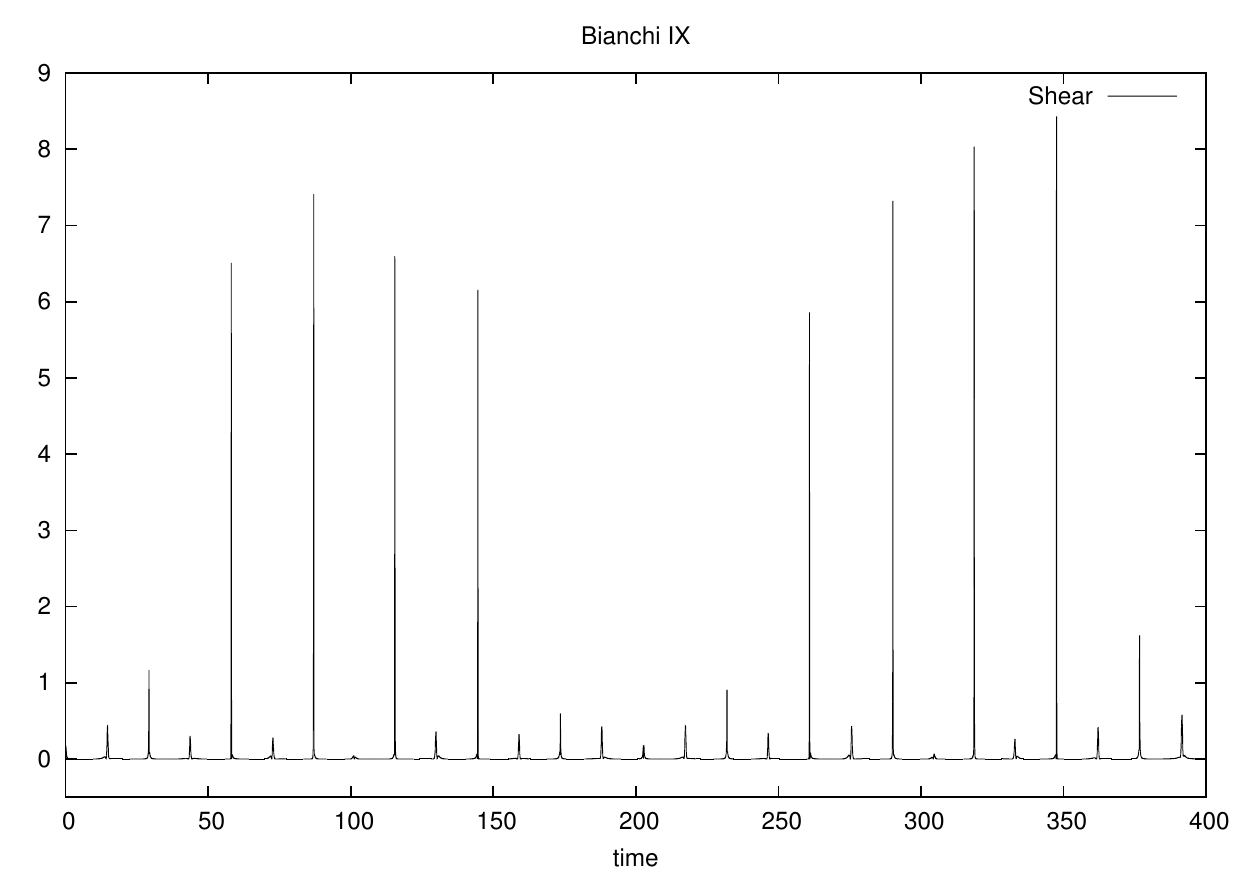}
        \end{subfigure}
     \begin{subfigure}
                \centering
                \includegraphics[width=0.47\textwidth ]{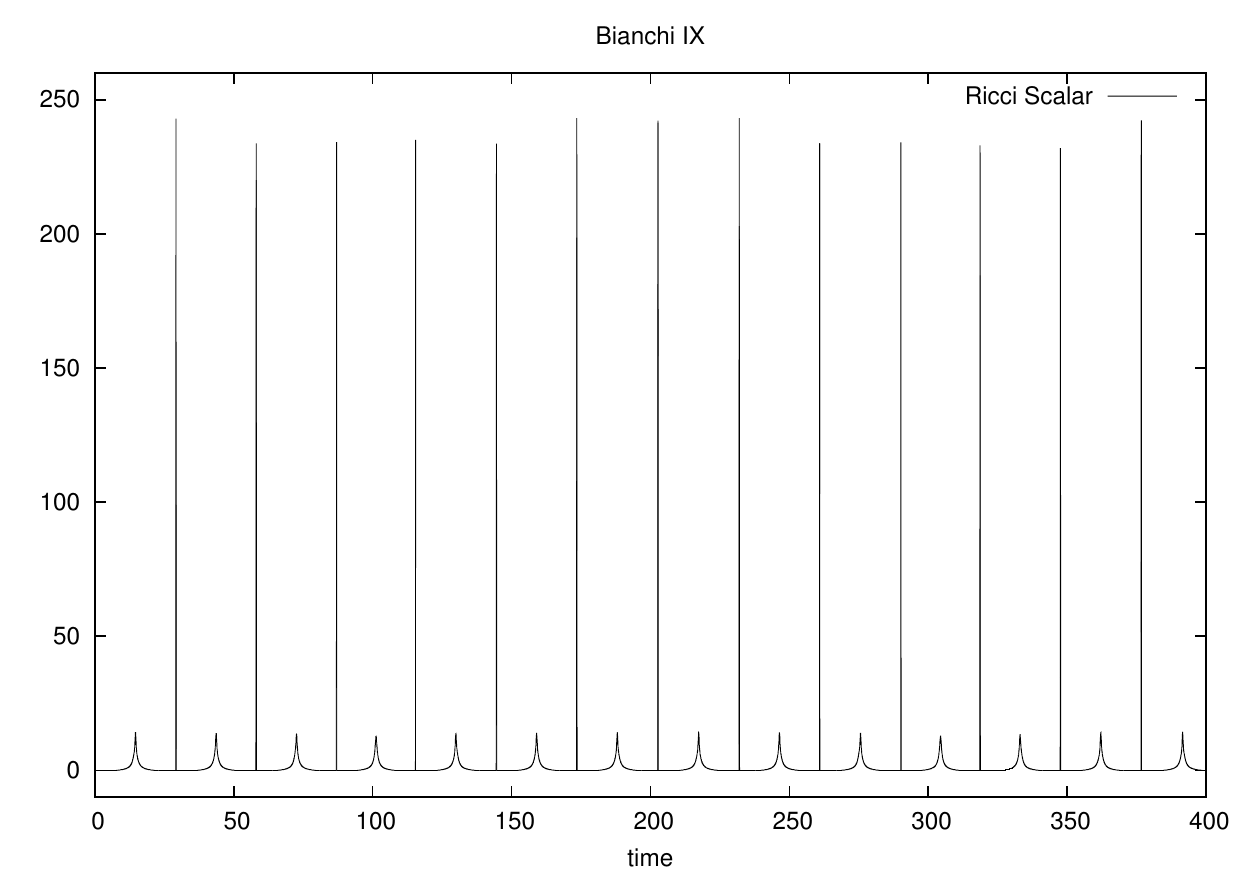}
        \end{subfigure}
\caption{Evolution of shear $\sigma^2$ and Ricci scalar $R$ for Bianchi IX. The initial conditions at $t=0$ are:
$\bar\mu_ic_i=\pi/2$, $p_1=90$, $p_2=80$ and $p_3=100$. }
\label{fig-b9}
\end{figure}

%%%%%%%%%%%%%%%%%%%%%%%%%%%%%%%%%%%%%%%%%%%%%%%%%%%%%%%%%
\begin{figure}
\includegraphics[width=0.5\textwidth]{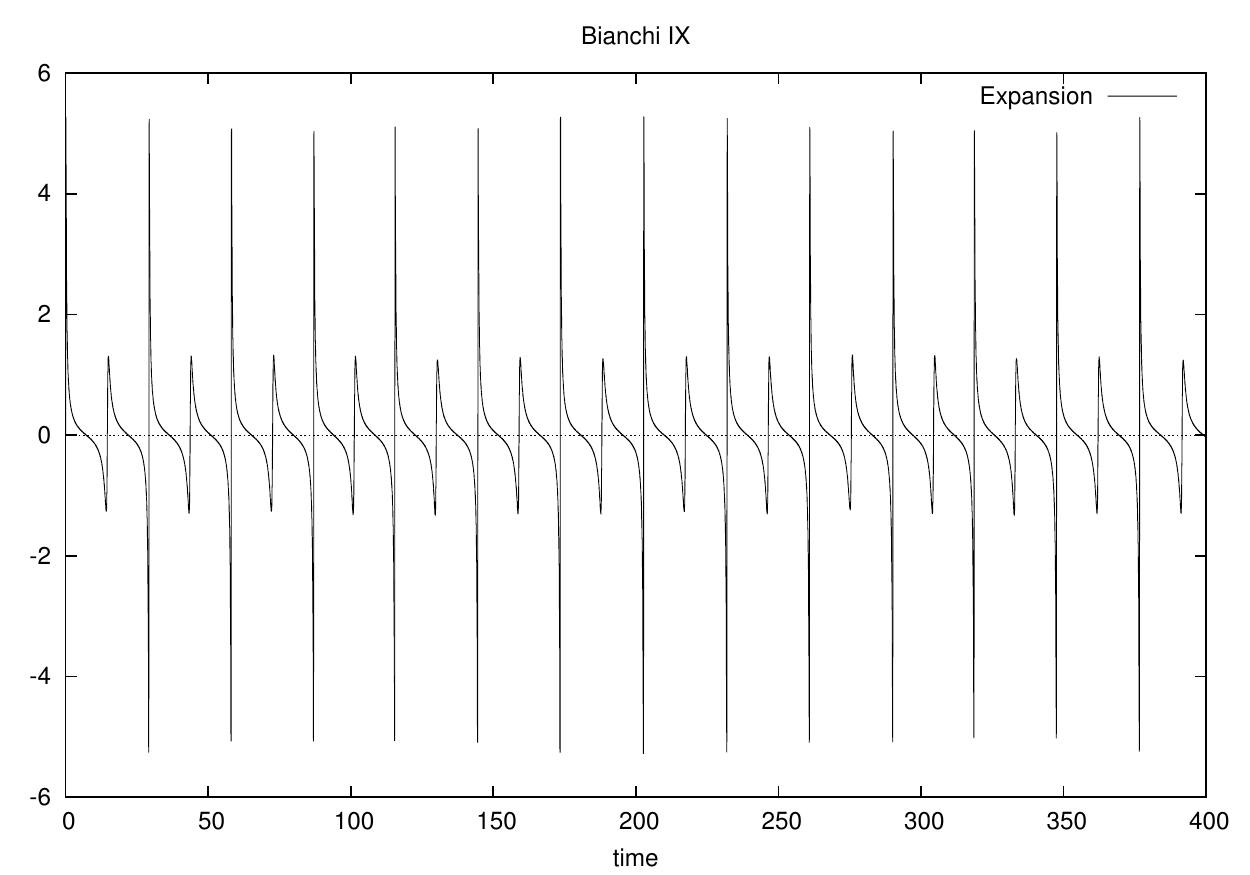}
\vspace{-11pt}
\caption{Evolution of expansion $\theta$ for Bianchi IX.}
\label{fig-b9-2}
\end{figure}

\item Can we have different kind of bounces, say, dominated by shear $\sigma^2$? Yes, but only in Bianchi II and IX. in Bianchi I the dynamical
contribution from matter is always bigger than the one from the shear, even in the solution which reaches the maximal shear at the bounce \cite{CM-bianchi2}.

\item Are geometric scalars $\theta, \sigma^2$ and $\rho$ absolutely bounded?
In the flat isotropic model all the solutions to the effective equations
have a maximal density equal to the critical density,
$\rho_{\rm crit} = 3/8\pi G \gamma^2\lambda^2$,
and a maximal expansion ($\theta^2_{\rm max} = 6\pi G \rho_{\rm crit} = 3/(2\gamma\lambda)$)
when $\rho=\rho_{\rm crit}/2$.
For the FRW $k=1$ model, every solution has its maximum density but in general the density is not absolutely bounded.
In the effective theory which comes from connection based quantization, expansion can tend to infinity.
For the other case, expansion has the same bound as in the flat FRW model. However, by adding some more corrections
from inverse triad term, Eqs. (\ref{c-k1}, \ref{c-k11}), one can show that actually in both effective theories the density and
the expansion have finite values.
For Bianchi I, in all the solutions $\rho$ and $\theta$ are upperly bounded by its values in the isotropic case
and $\sigma^2$ is bounded by $\sigma^2_{\max} = 10.125/(3\gamma^2\lambda^2)$ \cite{singh-gupt}.
For Bianchi II, $\theta, \sigma^2$ and $\rho$ are also bounded, but for larger values than the ones in Bianchi I,
i.e., there are solutions where the matter density is larger than the
critical density. With point-like and cigar-like classical singularities \cite{CM-bianchi2},
the density can achieve the maximal value ($\rho \approx 0.54\rho_p $) as a consequence of the shear
being zero at the bounce and curvature different from zero.
For Bianchi IX the behavior is the same as in the closed FRW; if the inverse triad corrections are not included, then
the geometric scalars are not absolutely bounded. But if the inverse triad corrections are included then,
on each solution, the geometric scalars
are bounded but there is not an absolute bound for all the solutions \cite{CKM,singh-gupt}.
\end{itemize}

There are others results that are important to mention,

\begin{itemize}
\item Classical and effective solutions are equal far away from the bounce.

\item Bianchi I, II and therefore the isotropic case $k$=0 are limiting cases of Bianchi IX, but they are not contained within Bianchi IX.
While the isotropic FRW $k$=1 is contained within Bianchi IX only if the inverse triad corrections are not included, when they are included then the $k$=1 universe is a limiting case, like the $k$=0 universe.

\item A set of quantities that are very useful
are the Kasner exponents (in classical Bianchi I, the scales factors are $a_i=t^{k_i}$, where $k_i$ are the Kasner
exponents), because they can be used to determine which kind of solution one obtains.
The Kasner exponents tell us about the Bianchi I transitions \cite{singh-gupt2} (if they exist) and particularly
in Bianchi IX, they are used to study the BKL behavior in the vacuum case.

\item Some important solutions are locally rotational symmetric.
This means that, at each point the spacetime is invariant under rotation about a preferred spacelike axis,
for example, if $a_2=a_3$.
These solutions are such as the one with maximal density and the vacuum limit (where all the dynamical contribution
come from the anisotropies).
\end{itemize}

%%%%%%%%%%%%%%%%%%%%%%%%%%%%%%%%%%%%%%%%%%%%%%%%%%%%
\section{Conclusions}

One of the main issues that a quantum theory of gravity is expected to address is that of
singularity resolution. Loop quantum cosmology has provided a complete description of the quantum dynamics
in the case of isotropic cosmological models and singularity resolution has been shown to be generic.
A pressing question is whether these results can be generalized to anisotropic models. %with and without spatial curvature.
In this case we lack a complete quantum theory, but one can rely on the existence of an effective description, capturing the main (loop)
quantum geometric features. In this contribution we have described the main features of such effective solutions, for a variety of anisotropic
cosmological models, with and without spatial curvature.
As we have seen, singularities seem to be generically resolved as the time evolution of geometrical scalars is well behaved past the
would-be classical singularity. The big bang is replaced by a big bounce.

We have also studied the behavior of a modified potential for the vacuum Bianchi IX model when quantum
effects become important. We showed that the potential wall does not disappear and we have, potentialy,
chaotic behavior near the classical singularity. However, if the weak energy condition holds and, if we start
from large volumes and evolve the equations into small volumes, there will be a lower bound for volume within
region B (Fig.\ref{dens}, right), and one does not reach region A (connected to zero volume).
Since there are no large anisotropies near the smallest allowed volume, the solutions will {\it not} exhibit
chaotic behavior.

With the study of these anisotropic models,
a question that still arises is whether this bouncing non-singular behavior is generic for inhomogeneous configurations.
That is, are we a step forward toward generic quantum singularity resolution?

\section*{Acknowledgments}
\noindent
 This work was in part supported by DGAPA-UNAM IN103610 grant, by NSF
PHY0854743, and the Eberly Research Funds of Penn State.

%%%%%%%%%%%%%%%%%%%%%%%%%%%%%%%%%%%%%%%%%%%%%%%%%%%%%%

\end{document}